\documentclass[a4paper,conference]{IEEEtran}
\usepackage[left=1.57cm,right=1.57cm,top=0.95cm,bottom=2.54cm]{geometry}

\usepackage{graphicx}
\usepackage{algorithm}
\usepackage{algpseudocode}
\usepackage{setspace}
\usepackage[font=scriptsize]{subcaption}
\usepackage[font=footnotesize]{caption}
\usepackage{mathtools}
\usepackage{amsthm}
\usepackage{cite}
\usepackage{amssymb}
\usepackage[capitalize]{cleveref}
\crefname{section}{Sec.}{Secs.}

\usepackage[nowarn,acronyms,nonumberlist,nopostdot,nomain,nogroupskip]{glossaries}
\usepackage{xcolor}

\makeatletter
\newcommand{\LeftEqNo}{\let\veqno\@@leqno}
\makeatother

\newacronym{3gpp}{3GPP}{3rd Generation Partnership Project}
\newacronym{5g}{5G}{5\textsuperscript{th} Generation}
\newacronym{5gc}{5GC}{5G Core}
\newacronym{bs}{BS}{Base Station}
\newacronym{abft}{A-BFT}{Association-BeamForming Training}
\newacronym[firstplural=Access Categories (ACs)]{ac}{AC}{Access Category}
\newacronym{adc}{ADC}{Analog to Digital Converter}
\newacronym{addts}{ADDTS}{Add Traffic Stream}
\newacronym{afbw}{AFBW}{Average Fading Bandwidth}
\newacronym{aid}{AID}{Association Identifier}
\newacronym{aimd}{AIMD}{Additive Increase Multiplicative Decrease}
\newacronym{am}{AM}{Acknowledged Mode}
\newacronym{amc}{AMC}{Adaptive Modulation and Coding}
\newacronym{ampdu}{A-MPDU}{MAC Protocol Data Unit Aggregation}
\newacronym{aoa}{AoA}{Angle of Arrival}
\newacronym{aod}{AoD}{Angle of Departure}
\newacronym{ap}{AP}{Access Point}
\newacronym{app}{APP}{Application Layer}
\newacronym{aqm}{AQM}{Active Queue Management}
\newacronym{ar}{AR}{Augmented Reality}
\newacronym{ati}{ATI}{Announcement Transmission Interval}
\newacronym{awgn}{AGWN}{Additive White Gaussian Noise}
\newacronym{awv}{AWV}{Antenna Weight Vector}
\newacronym{balia}{BALIA}{Balanced Link Adaptation}
\newacronym{bdp}{BDP}{Bandwidth-Delay Product}
\newacronym{bf}{BF}{Beamforming}
\newacronym{bhi}{BHI}{Beacon Header Interval}
\newacronym{bi}{BI}{Beacon Interval}
\newacronym{brp}{BRP}{Beam Refinement Protocol}
\newacronym{bss}{BSS}{Basic Service Set}
\newacronym{bti}{BTI}{Beacon Transmission Interval}
\newacronym{cad}{CAD}{Computer-aided Design}
\newacronym{cbap}{CBAP}{Contention-Based Access Period}
\newacronym{cbr}{CBR}{Constant Bitrate}
\newacronym{cc}{CC}{Congestion Control}
\newacronym{cdf}{CDF}{Cumulative Distribution Function}
\newacronym{cir}{CIR}{Channel Impulse Response}
\newacronym{cn}{CN}{Core Network}
\newacronym{cp}{CP}{Control Plane}
\newacronym{cqi}{CQI}{Channel Quality Indicator}
\newacronym{crs}{CRS}{Cell Reference Signal}
\newacronym{csirs}{CSI-RS}{Channel State Information - Reference Signal}
\newacronym{csmaca}{CSMA/CA}{Carrier Sense Multiple Access with Collision Avoidance}
\newacronym{cts}{CTS}{Clear to Send}
\newacronym{dc}{DC}{Dual Connectivity}
\newacronym{dce}{DCE}{Direct Code Execution}
\newacronym{dcf}{DCF}{Distributed Coordination Function}
\newacronym{dci}{DCI}{Downlink Control Information}
\newacronym{delts}{DELTS}{Delete Traffic Stream}
\newacronym{dl}{DL}{Downlink}
\newacronym{dmg}{DMG}{Directional Multi-Gigabit}
\newacronym{dmr}{DMR}{Deadline Miss Ratio}
\newacronym{dmrs}{DMRS}{DeModulation Reference Signal}
\newacronym{dti}{DTI}{Data Transmission Interval}
\newacronym{e2e}{E2E}{End-to-End}
\newacronym{ecn}{ECN}{Explicit Congestion Notification}
\newacronym{edca}{EDCA}{Enhanced Distributed Channel Access}
\newacronym{edf}{EDF}{Earliest Deadline First}
\newacronym{enb}{eNB}{evolved Node Base}
\newacronym{endc}{EN-DC}{E-UTRAN-\gls{nr} \gls{dc}}
\newacronym{epc}{EPC}{Evolved Packet Core}
\newacronym{es}{ES}{Edge Server}
\newacronym{ese}{ESE}{Extended Schedule Element}
\newacronym{fdd}{FDD}{Frequency Division Duplexing}
\newacronym{fdma}{FDMA}{Frequency Division Multiple Access}
\newacronym{fov}{FoV}{Field-of-View}
\newacronym{fs}{FS}{Fast Switching}
\newacronym{ftp}{FTP}{File Transfer Protocol}
\newacronym{gnb}{gNB}{Next Generation Node Base}
\newacronym{harq}{HARQ}{Hybrid Automatic Repeat reQuest}
\newacronym{hetnet}{HetNet}{Heterogeneous Network}
\newacronym{hh}{HH}{Hard Handover}
\newacronym{hol}{HOL}{Head-of-Line}
\newacronym{hqf}{HQF}{Highest-quality-first}
\newacronym{ia}{IA}{Initial Access}
\newacronym{iab}{IAB}{Integrated Access and Backhaul}
\newacronym{ibss}{IBSS}{Independent Basic Service Set}
\newacronym{id}{ID}{Identifier}
\newacronym{imt}{IMT}{International Mobile Telecommunication}
\newacronym{inr}{INR}{Interference to Noise Ratio}
\newacronym{iot}{IoT}{Internet of Things}
\newacronym{ipa}{IPA}{Inter-Packet Arrival}
\newacronym{ism}{ISM}{Industrial, Scientific, and Medical}
\newacronym{kpi}{KPI}{Key Performance Indicator}
\newacronym{lcf}{LCF}{Level Crossing Frequency}
\newacronym{lcr}{LCR}{Level Crossing Rate}
\newacronym{los}{LoS}{Line-of-Sight}
\newacronym{lp}{LP}{Low Power}
\newacronym{lte}{LTE}{Long Term Evolution}
\newacronym{m2m}{M2M}{Machine to Machine}
\newacronym{mac}{MAC}{Medium Access Control}
\newacronym{mc}{MC}{Multi-Connectivity}
\newacronym{mcs}{MCS}{Modulation and Coding Scheme}
\newacronym{mec}{MEC}{Mobile Edge Cloud}
\newacronym{mi}{MI}{Mutual Information}
\newacronym{mib}{MIB}{Master Information Block}
\newacronym{mimo}{MIMO}{Multiple Input, Multiple Output}
\newacronym{mumimo}{MU-MIMO}{Multi-User Multiple Input, Multiple Output}
\newacronym{ml}{ML}{Machine Learning}
\newacronym{mlr}{MLR}{Maximum-local-rate}
\newacronym[plural=\gls{mme}s,firstplural=Mobility Management Entities (MMEs)]{mme}{MME}{Mobility Management Entity}
\newacronym{mmw}{mmW}{Millimeter Wave}
\newacronym{moi}{MoI}{Method of Images}
\newacronym{mpc}{MPC}{Multi Path Component}
\newacronym{mptcp}{MPTCP}{Multipath TCP}
\newacronym{mr}{MR}{Maximum Rate}
\newacronym{mrdc}{MR-DC}{Multi \gls{rat} \gls{dc}}
\newacronym{mss}{MSS}{Maximum Segment Size}
\newacronym{mtd}{MTD}{Machine-Type Device}
\newacronym{mtu}{MTU}{Maximum Transmission Unit}
\newacronym{nav}{NAV}{Network Allocation Vector}
\newacronym{ncbr}{NCBR}{Non-Constant Bitrate}
\newacronym{nfv}{NFV}{Network Function Virtualization}
\newacronym{nlos}{NLoS}{Non-Line-of-Sight}
\newacronym{nr}{NR}{New Radio}
\newacronym{nrmse}{NRMSE}{Normalized Root Mean Square Error}
\newacronym{ns3}{ns-3}{Network Simulator 3}
\newacronym{nsa}{NSA}{Non Stand Alone}
\newacronym{o2i}{O2I}{Outdoor-to-Indoor}
\newacronym{ofdm}{OFDM}{Orthogonal Frequency Division Multiplexing}
\newacronym{pa}{PA}{Position-aware}
\newacronym{pan}{PAN}{Personal Area Network}
\newacronym{pbch}{PBCH}{Physical Broadcast Channel}
\newacronym{pbss}{PBSS}{Personal Basic Service Set}
\newacronym{pcp}{PCP}{\gls{pbss} Central Point}
\newacronym{pcpap}{PCP/AP}{\acrlong{pcp}/\acrlong{ap}}
\newacronym{pdcch}{PDCCH}{Physical Downlonk Control Channel}
\newacronym{pdcp}{PDCP}{Packet Data Convergence Protocol}
\newacronym{pdsch}{PDSCH}{Physical Downlink Shared Channel}
\newacronym{pdu}{PDU}{Packet Data Unit}
\newacronym{pf}{PF}{Proportional Fair}
\newacronym{pgw}{PGW}{Packet Gateway}
\newacronym{phy}{PHY}{Physical Layer}
\newacronym{ppp}{PPP}{Poisson Point Process}
\newacronym{prb}{PRB}{Physical Resource Block}
\newacronym{pss}{PSS}{Primary Synchronization Signal}
\newacronym{pucch}{PUCCH}{Physical Uplink Control Channel}
\newacronym{pusch}{PUSCH}{Physical Uplink Shared Channel}
\newacronym{qd}{QD}{Quasi Deterministic}
\newacronym{qos}{QoS}{Quality of Service}
\newacronym{rach}{RACH}{Random Access Channel}
\newacronym{ran}{RAN}{Radio Access Network}
\newacronym[firstplural=Radio Access Technologies (RATs)]{rat}{RAT}{Radio Access Technology}
\newacronym{red}{RED}{Random Early Detection}
\newacronym{rf}{RF}{Radio Frequency}
\newacronym{rl}{RL}{Reinforcement Learning}
\newacronym{rlc}{RLC}{Radio Link Control}
\newacronym{rlf}{RLF}{Radio Link Failure}
\newacronym{rr}{RR}{Round Robin}
\newacronym{rrc}{RRC}{Radio Resource Control}
\newacronym{rrm}{RRM}{Radio Resource Management}
\newacronym{rs}{RS}{Remote Server}
\newacronym{rsrp}{RSRP}{Reference Signal Received Power}
\newacronym{rsrq}{RSRQ}{Reference Signal Received Quality}
\newacronym{rss}{RSS}{Received Signal Strength}
\newacronym{rssi}{RSSI}{Received Signal Strength Indicator}
\newacronym{rt}{RT}{Ray Tracer}
\newacronym{rts}{RTS}{Request to Send}
\newacronym{rtt}{RTT}{Round Trip Time}
\newacronym{rw}{RW}{Receive Window}
\newacronym{rx}{RX}{Receiver}
\newacronym{sa}{SA}{standalone}
\newacronym{sack}{SACK}{Selective Acknowledgment}
\newacronym{sap}{SAP}{Service Access Point}
\newacronym{sc}{SC}{Single Carrier}
\newacronym{sch}{SCH}{Secondary Cell Handover}
\newacronym{scm}{SCM}{Spatial Channel Model}
\newacronym{scoot}{SCOOT}{Split Cycle Offset Optimization Technique}
\newacronym{sdma}{SDMA}{Spatial Division Multiple Access}
\newacronym{sdr}{SDR}{Software Defined Radio}
\newacronym{si}{SI}{Study Item}
\newacronym{sib}{SIB}{Secondary Information Block}
\newacronym{sinr}{SINR}{Signal-to-Interference-plus-Noise Ratio}
\newacronym{sir}{SIR}{Signal-to-Interference Ratio}
\newacronym{sls}{SLS}{Sector-Level Sweep}
\newacronym{sm}{SM}{Saturation Mode}
\newacronym{snr}{SNR}{Signal-to-Noise Ratio}
\newacronym{son}{SON}{Self-Organizing Network}
\newacronym{sp}{SP}{Service Period}
\newacronym{spr}{SPR}{Service Period Request}
\newacronym{srs}{SRS}{Sounding Reference Signal}
\newacronym{ss}{SS}{Synchronization Signal}
\newacronym{sss}{SSS}{Secondary Synchronization Signal}
\newacronym{ssw}{SSW}{Sector Sweep}
\newacronym{sta}{STA}{Station}
\newacronym{stb}{STB}{Set Top Box}
\newacronym{tb}{TB}{Transport Block}
\newacronym{tcp}{TCP}{Transmission Control Protocol}
\newacronym{tdd}{TDD}{Time Division Duplexing}
\newacronym{tdma}{TDMA}{Time Division Multiple Access}
\newacronym{tfl}{TfL}{Transport for London}
\newacronym{tgad}{TGad}{Task Group ad}
\newacronym{tgay}{TGay}{Task Group ay}
\newacronym{tsconst}{TSCONST}{Traffic Scheduling Constraint}
\newacronym{tm}{TM}{Transparent Mode}
\newacronym{trp}{TRP}{Transmitter Receiver Pair}
\newacronym{ts}{TS}{Traffic Stream}
\newacronym{tspec}{TSPEC}{Traffic Specification}
\newacronym{tti}{TTI}{Transmission Time Interval}
\newacronym{ttt}{TTT}{Time-to-Trigger}
\newacronym{tx}{TX}{Transmitter}
\newacronym[firstplural=Transmission Opportunities (TXOPs)]{txop}{TXOP}{Transmission Opportunity}
\newacronym{udp}{UDP}{User Datagram Protocol}
\newacronym{ue}{UE}{User Equipment}
\newacronym{ul}{UL}{Uplink}
\newacronym{um}{UM}{Unacknowledged Mode}
\newacronym{uma}{UMa}{Urban Macro}
\newacronym{uml}{UML}{Unified Modeling Language}
\newacronym{utc}{UTC}{Urban Traffic Control}
\newacronym{vbr}{VBR}{Variable Bit Rate}
\newacronym{vm}{VM}{Virtual Machine}
\newacronym{vr}{VR}{Virtual Reality}
\newacronym{wbf}{WBF}{Wired Bias Function}
\newacronym{wf}{WF}{Wired-first}
\newacronym{wifi}{Wi-Fi}{Wireless Fidelity}
\newacronym{wigig}{WiGig}{Wireless Gigabit}
\newacronym{wlan}{WLAN}{Wireless Local Area Network}
\newacronym{ber}{BER}{Bit Error Rate}
\newacronym{arf}{ARF}{Auto Rate Fallback}
\newacronym{semm}{SEMM}{SPCA-EDCA Mixed Mode}
\newacronym{ppdu}{PPDU}{PHY Protocol Data Unit}
\makeglossaries
\glsdisablehyper

\IEEEoverridecommandlockouts
\usepackage{tikz}
\newcommand\copyrightnotice{%
\begin{tikzpicture}[remember picture,overlay]
\node[anchor=south,yshift=10pt] at (current page.south) {\fbox{\parbox{\dimexpr\textwidth-\fboxsep-\fboxrule\relax}{
\footnotesize \textcopyright 2020 IEEE. Personal use of this material is permitted.
Permission from IEEE must be obtained for all other uses, in any current or future media,
including reprinting/republishing this material for advertising or promotional purposes,
creating new collective works, for resale or redistribution to servers or lists,
or reuse of any copyrighted component of this work in other works.}}};
\end{tikzpicture}
}

\begin{document}


\title{
    The challenges of Scheduling and\\Resource Allocation in IEEE 802.11ad/ay
}

\author{
    \IEEEauthorblockN{Salman Mohebi, Mattia Lecci, Andrea Zanella, Michele Zorzi}
    \IEEEauthorblockA{Department of Information Engineering, University of Padova, Italy\\
                      E-mails: \texttt{\{name.surname\}@dei.unipd.it}}
    \thanks{This work was partially supported by NIST under Award No. 60NANB19D122.
    Mattia Lecci's activities were also supported by \textit{Fondazione CaRiPaRo} under the grant ``Dottorati di Ricerca 2018.''
    Salman Mohebi's activities were supported by the EU H2020 project ``WindMill,'' under the MSC Grant Agreement No.~813999}
}

\maketitle
\copyrightnotice

\begin{abstract}
The IEEE 802.11ad \acrshort{wifi} amendment enables short-range multi-gigabit communications in the unlicensed 60~GHz spectrum, unlocking new interesting applications such as wireless Augmented and Virtual Reality.
The characteristics of the \gls{mmw} band and directional communications allow increasing the system throughput by scheduling pairs of nodes with low cross-interfering channels in the same time-frequency slot.
On the other hand, this requires significantly more signaling overhead.
Furthermore, IEEE 802.11ad introduces a hybrid \acrshort{mac} characterized by two different channel access mechanisms: contention-based and contention-free access periods.
The coexistence of both access period types and the directionality typical of \gls{mmw} increase the channel access and scheduling complexity in IEEE 802.11ad compared to previous \acrshort{wifi} versions.
Hence, to provide the \acrlong{qos} performance required by demanding applications, a proper resource scheduling mechanism that takes into account both directional communications and the newly added features of this \acrshort{wifi} amendment is needed.
In this paper, we present a brief but comprehensive review of the open problems and challenges associated with channel access in IEEE 802.11ad and propose a workflow to tackle them via both heuristic and learning-based methods.
\end{abstract}


\begin{tikzpicture}[remember picture,overlay]
\node[anchor=north,yshift=2pt] at (current page.north) {\fbox{\parbox{\dimexpr\textwidth-\fboxsep-\fboxrule\relax}{
\centering\footnotesize This paper has been presented at IEEE MedComNet 2020. \textcopyright 2020 IEEE.\\
Please cite it as: S. Mohebi, M. Lecci, A. Zanella, M. Zorzi, ``The challenges of Scheduling and
Resource Allocation in IEEE 802.11ad/ay,'' in 18th Mediterranean Communication and Computer Networking Conference (MedComNet),
Arona, Italy, Jun. 2020.}}};
\end{tikzpicture}

\section{Introduction}
\label{sec:introduction}
\glsresetall
\glsunset{wifi}

\Gls{wifi} is nowadays present in many devices and is common in households, offices, public institutions, and transportation.
Over more than 20~years, many amendments have been made to the original standard, updating both the \gls{phy} and \gls{mac} layers to provide higher bit-rate, robustness, and \gls{qos}.

As users keep asking for higher data-rates, the current deployments struggle to keep up with the demand.
One key enabler for gigabit-class communications is the use of the \gls{mmw} band, which loosely refers to the portion of the electromagnetic spectrum with frequencies higher than 6~GHz.
In this frequency range, the amount of available bandwidth is significantly larger than that of the legacy sub-6~GHz counterpart, allowing unprecedented transfer speeds.

As the research started to mature, the \gls{wifi} Alliance introduced in 2012 the IEEE 802.11ad amendment~\cite{standard802.11_2016}, standardizing communication in the 60~GHz \gls{ism} unlicensed band, offering data-rates up to 6.75~Gbps.
As a follow-up, its successor IEEE 802.11ay is planned to be standardized by the end of 2020~\cite{tgayWebsite}, introducing technologies such as \gls{mumimo}, channel bonding, higher-order modulation, and thus even higher speeds.
Such extreme data-rates make it possible to unlock new applications, such as wireless office docking, 8K Ultra High Definition video transfer, wireless \gls{ar} and \gls{vr}, mobile front-hauling and offloading, etc.~\cite{tgayUsageModel}.

On the downside, given the higher carrier frequency, \gls{mmw} transmission suffers from an increased propagation loss, as well as deeper diffraction shadows, and higher penetration and reflection losses, making communication more difficult and less stable.

On the other hand, these characteristics allow for extreme spatial reuse, e.g., transmissions in different rooms will hardly interfere with each other unlike in legacy \gls{wifi}.
Moreover, the short wavelength makes it possible to use antenna arrays with tens of elements packed in a small area, making it is possible to counteract the increased path loss by focusing the radiated power into directive \emph{beams}, thus increasing the overall antenna gain.
While this further reduces interference even where users share the same area and improves spatial reuse, it also creates the problem of directional deafness, worsens the hidden node problem, and makes mobility more complex to handle.

To meet the strict \gls{qos} requirements of some new applications and partially alleviate the hidden node problem, the standard provides the possibility to transmit data in reserved contention-free periods, that coexist with contention-based access periods, very similar to the legacy \gls{csmaca} channel access mechanism, and the hybrid allocation can be flexible enough to support the coexistence of traffic with vastly different \gls{qos} requirements.

In this paper, we present some of the challenges related to the scheduling of IEEE 802.11ad/ay devices in realistic scenarios, with the main focus on the already-standardized IEEE 802.11ad.
Furthermore, we discuss some pre-existing works and propose some research directions.

In particular, in \cref{sec:ieee_802_11ad_overview} the main characteristics of IEEE 802.11ad will be described.
\Cref{sec:scheduling_in_ieee_802_11ad} will briefly discuss the literature on channel access and scheduling.
\Cref{sec:future_research} will showcase our research plan, and finally \cref{sec:conclusions} will draw the conclusions.

\section{IEEE 802.11ad Overview}
\label{sec:ieee_802_11ad_overview}

To introduce the main concepts and nomenclature of IEEE 802.11ad, in this section we provide a short summary of the standard~\cite{standard802.11_2016}, while referring to other sources for more details~\cite{11ad_magazine}.

\begin{figure}[tp]
    \centering
    \includegraphics[width=.9\columnwidth]{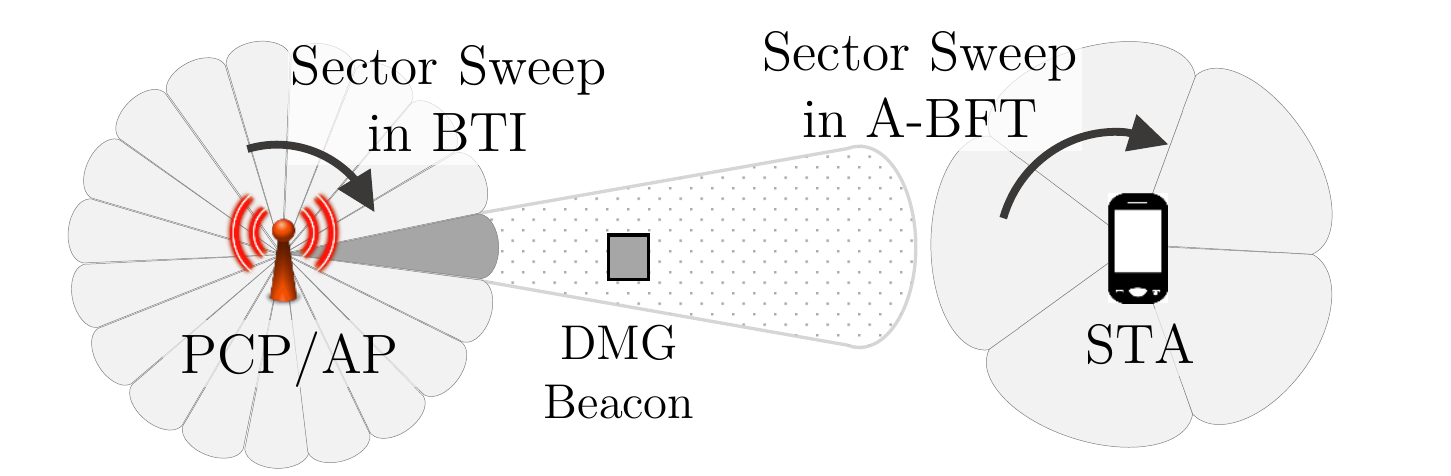}
    \caption{Graphical representation of sector structure in IEEE 802.11ad.}
    \label{fig:model}
\end{figure}

Being a \gls{mmw}-based standard, directional communication with all the added overhead and the possibility of spatially multiplexing users are included in the amendment.
To simplify beam management, both the \gls{pcpap} and the \glspl{sta} divide their surrounding space into sectors as shown in \cref{fig:model}.
\Glspl{sta} and \gls{pcpap} will then need to keep beam alignment, which increases the signaling overhead.

\begin{figure}[tp]
    \centering
    \includegraphics[width=.9\columnwidth]{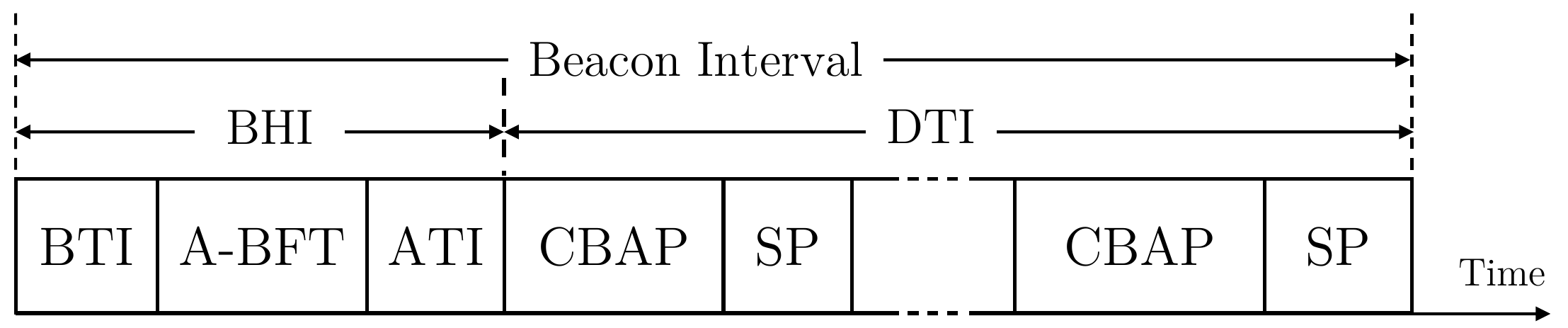}
    \caption{Representation of a \acrfull{bi}.}
    \label{fig:interval}
\end{figure}

\cref{fig:interval} shows that in IEEE 802.11ad time is divided in \glspl{bi} of about 100~ms.
Each \gls{bi} is further divided into \gls{bhi} and \gls{dti}, briefly described in the following sections.

\subsection{Beacon Header Interval}
\label{sub:bhi}

The \gls{pcpap} does most of the managing, such as beaconing, beamforming training, and scheduling, during the \gls{bhi}.
This period can last hundreds of microseconds up to a few milliseconds, and is further divided into three subintervals: \gls{bti}, \gls{abft}, and \gls{ati}.

The \gls{bti} is used to send \gls{dmg} Beacons to announce the network, give the basic synchronization and \gls{bi} structure information, start the beamforming training with new stations, and, if needed, do some basic traffic management.
Beacons are sent over the different sectors, covering all possible directions to maximize coverage for untrained \glspl{sta}.

After receiving a \gls{dmg} Beacon during the \gls{bti}, new \glspl{sta} can use the \gls{abft} to complete the basic beamforming training by sending \gls{ssw} frames in different sectors.
Beam alignment is completed once the \gls{pcpap} responds with an \gls{ssw} Feedback.

Finally, advanced scheduling mechanisms setup and further network management can be done during the optional \gls{ati}.

\subsection{Data Transmission Interval}
\label{sub:dti}

The \gls{dti} is mainly used for the actual data transmission, but it can also be used to improve communication links and for further scheduling.
The \gls{dti} comprises \acrfullpl{cbap} and \acrfullpl{sp}, which can appear in arbitrary combinations and are scheduled during the \gls{bhi}.

Transmission in \gls{cbap} follows the rule of \gls{edca}, slightly modified to account for directional transmission, in which \glspl{sta} compete with each other in order to transmit their data.

Instead, \glspl{sp} are scheduled contention-free intervals that are dedicated to exclusive transmission between a pair of \glspl{sta}\footnote{A \gls{pcpap} also \textit{contains} a \gls{sta}, i.e., \textit{a logical entity that is a singly addressable instance of a \gls{mac} and \gls{phy} interface to the wireless medium}~\cite{standard802.11_2016}.} to guarantee \gls{qos}.
The standard also allows for spatial sharing, meaning that multiple pairs of \glspl{sta} with low cross-interference can be scheduled in the same \glspl{sp}.
This, however, comes at the cost of increased overhead since interference measurements must periodically take place.

\section{Scheduling in IEEE 802.11ad}
\label{sec:scheduling_in_ieee_802_11ad}

IEEE 802.11ad allows for great flexibility in the scheduling of radio resources, but we will hereby describe only some of these possibilities in their simplest form.

We want to stress the fact that, unlike in traditional contention-based medium access, scheduled \glspl{sp} guarantee \gls{qos}.
\Glspl{ac} introduced in 802.11e, in fact, only allow for stochastic traffic prioritization according to the DiffServ paradigm, which ceases to work in congested networks.
For this reason, allocated traffic is especially important for those applications with strict \gls{qos} constraints.
Instead, more realistic applications, such as data transfer or asynchronous bursty traffic, can simply rely on \gls{cbap}.

\begin{figure}[tp]
    \centering
    \includegraphics[width=.9 \columnwidth]{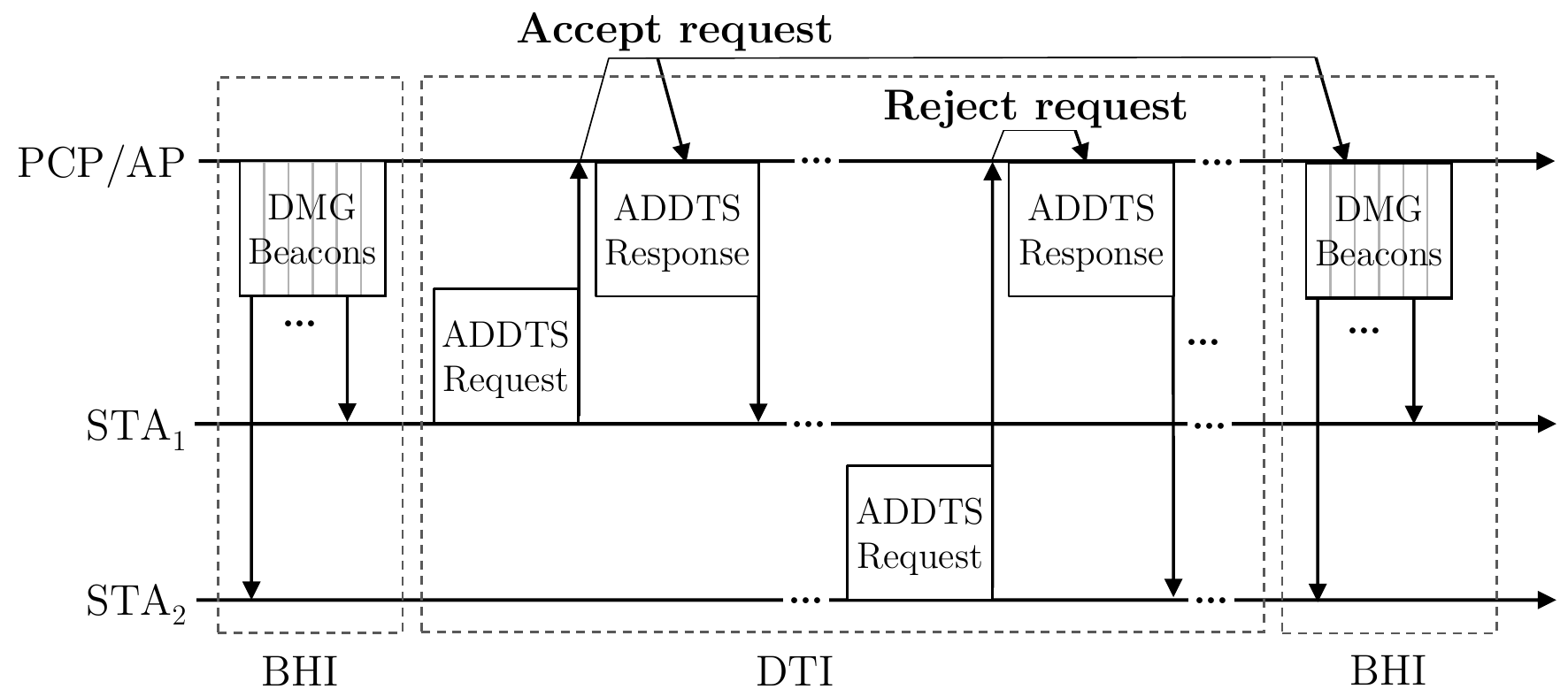}
    \caption{Representation of \acrshort{addts} scheduling in IEEE 802.11ad.}
    \label{fig:scheduling}
\end{figure}

As shown in \cref{fig:scheduling}, a \gls{sta} can set up an allocation by sending an \gls{addts} Request frame to the \gls{pcpap} during the \gls{dti} and embedding a \gls{dmg} \gls{tspec} element.
The \gls{dmg} \gls{tspec} element is created by the requesting \gls{sta} and comprises information such as the allocation period, and the minimum and maximum allocation duration.

Based on its admission policy, the \gls{pcpap} will either reject or accept the request, immediately notifying the requesting \gls{sta} via an \gls{addts} Response.
If accepted, the allocation is made effective by including it in the \gls{ese} transmitted in the next \gls{dmg} Beacons, which will contain details such as the effectively allocated period duration and the \gls{sp} start time.
In this way, \glspl{sta} not involved in the communication will not create interference and will be able to switch to power-saving mode.
Otherwise, the \gls{pcpap} can either reject or propose a change in the \gls{dmg} \gls{tspec}.
A \gls{sta} can later update the \gls{dmg} \gls{tspec} by sending another \gls{addts} Request with the updated element and follow again the same procedure.

Allocating the right duration to \glspl{sp} is clearly a trade-off between \gls{qos} traffic, which needs resources to fulfill the minimum requirements imposed by the application, and elastic traffic, which still needs resources even though with less stringent requirements.
Since resource availability, as well as channel quality, are time-varying, the standard supports \gls{sp} extension and truncation services, which let the stations keep transmitting and/or relinquishing the unused occupied channel.
Still, these features bring extra overhead and should thus be used carefully.

A mathematical model for preliminary allocation of \gls{sp} for \gls{vbr} flows is presented in~\cite{khorov2016mathematical}, which helps determine how to set the \gls{tspec} parameters to meet \gls{qos} requirements while minimizing the amount of allocated time.
Unfortunately, \glspl{sp} are assumed to be placed at the beginning of the \gls{dti}, which is not possible in general for applications with tight delay constraints.
For example, for virtual or augmented reality services, latencies should be below 20~ms to avoid motion sickness.

Other works in the literature consider different aspects of the \gls{dti}.
For example, \cite{rajan2017theoretical}~derives the theoretical maximum throughput for \glspl{cbap} when two-level MAC frame aggregation is used.
Beamforming is also considered in \cite{shokri2015beam}, which proposes a joint optimization of beamwidth selection and scheduling to maximize the effective network throughput, while other works, though not specifically concerning IEEE 802.11ad, deal with transmission scheduling for \gls{mmw} communications \cite{feng2017mmwave}.

\section{Future Research}
\label{sec:future_research}

In this section, we highlight some possible research directions.
In particular, in \cref{sub:available_research_tools} we describe the main tools that are currently available to study the subject.
Then, in \cref{sub:research_plan} we propose a possible research plan.

\subsection{Available Research Tools}
\label{sub:available_research_tools}

Although commercial devices supporting the IEEE 802.11ad standard are currently available, manufacturers do not provide tools to access low-level functionalities.
Ultimately, it is more flexible, timely, and cost-effective, although arguably less realistic, to simulate the behavior of such devices.

In particular, significant effort has already been done implementing the IEEE 802.11ad standard into \gls{ns3}~\cite{assasa2019enhancing}, a popular open-source network-level simulator.
The last release of the simulator also supports quasi-deterministic channel modeling based on ray-tracing, making simulations extremely accurate and realistic at the cost of a long preliminary channel generation phase, although some works already tried to improve this aspect~\cite{lecciIta2020}.
While the implementation already covers most parts of the standard, it is still missing the scheduling mechanisms necessary for this project.
The authors of~\cite{assasa2019enhancing} are also working on the implementation of the IEEE 802.11ay amendment~\cite{assasa2019ayProposal}, making their work even more valuable.

Historically, scheduling algorithms have been mainly based on heuristics, trying to balance performance and adaptiveness versus complexity.
In the last years, instead, the \gls{ml} revolution has brought many innovations also to the telecommunication field at all layers of the stack and, in particular, \gls{rl} is especially applicable to optimize or even replace legacy scheduling algorithms~\cite{azzino2020scheduling}.
Following the principle of \textit{self-driving networks}~\cite{feamster2018and}, \gls{ml} algorithms can learn from real on-line data and supersede manually-designed protocols, which are becoming increasingly complex.
OpenAI Gym is one of the most used \gls{rl} toolkits and has been adopted by all popular \gls{ml} frameworks.
Given their potential in many fields of networking and telecommunications in general, OpenAI Gym APIs have also been integrated into \gls{ns3}~\cite{gawlowicz2018ns3} with the name of \textit{ns3-gym}. 

With these powerful tools, it will be possible to further advance the state of the art, create a comprehensive performance evaluation of available algorithms and further improve upon them once the weak points are clearly identified.

\subsection{Research Plan}
\label{sub:research_plan}

One of our first goals is to extend the already existing IEEE 802.11ad \gls{ns3} module with the necessary mechanisms to make it properly support the hybrid channel access and advanced scheduling (see \cref{sub:available_research_tools}), and add the support to the ns3-gym framework.
A significant development effort will be put into the creation of a proper simulation environment, with particular attention to the computational complexity since a high data-rate simulation of just 10~s of simulated time may currently take one hour or more of run-time.
This makes the design, evaluation, and optimization of scheduling protocol a lengthy process, which may be even infeasible if \gls{rl} is involved since many training episodes are needed to learn even basic mechanisms.

Indeed, decisions such as admission policy, resource allocation, smart \gls{sp} truncation or extension, and spatial sharing are often difficult to accurately model in terms of trade-offs and usually comprise several tunable parameters.
However, if trained correctly, an \gls{rl} agent is often capable of learning extremely complex rules and optimize the network for different networking metrics (e.g., delay, jitter, throughput, fairness) even beyond complicated heuristics.

Resource allocation can be divided into two subproblems.
Specifically, \glspl{sta} have to translate information given by the application into \gls{dmg} \gls{tspec} elements and the \gls{pcpap} subsequently has to efficiently schedule the \gls{dti} especially considering the \gls{mcs} used.
Regarding the former, applications may not yield constant inter-packet arrival time (e.g., frame-rate drop in video applications) nor packet size (e.g., when compression is considered).
At the same time, transmission conditions may vary mainly due to environmental changes, mobility, or even blockage, thus increasing performance variability.
If \gls{qos} requirements are not met, the \gls{rl} agent of a \gls{sta} could thus update the \gls{tspec}.

On the other hand, the \gls{pcpap} has to allocate \glspl{sp} for a \gls{bi} based on the available resources.
Effective scheduling must take into account, in addition to network metrics, the possible evolution of the \gls{mcs} since the packet transmission time largely depends on it.
Given the significant differences in channel dynamics of IEEE 802.11ad with respect to sub-6~GHz \gls{wifi}, new ones can be proposed to account for the specific characteristics of the \gls{mmw} channel.
An \gls{rl} agent could thus jointly adapt the \gls{mcs} and perform scheduling to optimize the network performance by observing the evolution of both channel statistics and network traffic.

One way to overcome the problem of slow simulations is to quickly pre-train the \gls{rl} agent to make it learn at least simple decisions, such as understanding when a new request does not fit the available resources, avoid overlapping \glspl{sp} during scheduling, and avoid scheduling highly cross-interfering users with spatial sharing.
Thus, we plan to build a very simple and fast simulator that will only model relevant notions, e.g., basic channel and traffic modeling and the \gls{bi} structure defined in IEEE 802.11ad, but eliminating the fine details which make \gls{ns3} realistic but extremely slow.
In this way, the agent can learn very broadly which actions it should take and then fine-tune its behavior via more realistic simulations.
Then, to further decrease \gls{ns3} simulation run-time, a database of simulation results can be created and multiple agents can passively learn from it~\cite{munos2016_offPolicyRL} before fine-tuning their performance on ad hoc simulations.
Transfer learning will also be considered to speed up convergence to effective policies in different scenarios.

Another objective will be to understand the traffic behavior of target applications.
For example, it could be possible to acquire real-world traffic traces of \gls{ar}/\gls{vr} applications, characterizing and modeling their traffic patterns with focus on packet size, and variability of inter-packet arrival accounting for variable frame-rate statistics.
These models would ultimately be integrated with standardized scenarios~\cite{tgadEvaluationMethodology,tgayEvaluationMethodology} to further increase simulation realism.

Furthermore, understanding how the current state of the art performs in a realistic simulator will allow understanding the strong and weak points of the proposals in realistic settings.
From detailed studies, it will be possible to understand how the state of the art can be improved upon with heuristics or, when modeling becomes too complex or inaccurate, \gls{ml}-based approaches.

These results will then be easily transferred to the future IEEE 802.11ay standard, which will add further complexity on top of the already existing one, by introducing channel bonding and \gls{mumimo}.
Even more complex schedulers will then have to be designed, but starting from the solid ground of the proposed work further improvements will be possible.

\section{Conclusions}
\label{sec:conclusions}

In this paper we briefly described the main characteristics of IEEE 802.11ad, mainly focusing on the \gls{mac} layer and especially on the newly introduced scheduling mechanisms, allowing different types of traffic to coexist and potentially improving the performance of \gls{qos}-sensitive applications.
As shown in \cref{sec:scheduling_in_ieee_802_11ad}, some research has already been done in this direction but lacks a common and realistic testing ground, making it unclear whether the assumptions may hold.

Our future work will focus on proposing solutions for the many open problems described in \cref{sub:research_plan}.
Models and source code that will be considered of interest for the community will be published, making it possible to fairly compare results from different groups in a common and realistic simulation environment.



\end{document}